\documentclass[runningheads]{llncs}
\usepackage[T1]{fontenc}
\usepackage{graphicx}
\usepackage{hyperref}
\usepackage{color}

\definecolor{gray}{rgb}{0.4,0.4,0.4}
\definecolor{darkblue}{rgb}{0.0,0.0,0.6}
\definecolor{cyan}{rgb}{0.0,0.6,0.6}
\usepackage{listings}
\begin{document}
\title{Rich Interoperable Metadata for Cultural Heritage Projects at Jagiellonian University}
\titlerunning{Rich Interoperable Metadata for CH Projects at JU}
\author{Luiz do Valle Miranda\inst{1}\orcidID{0000-0003-1838-5693} \and
Krzysztof~Kutt\inst{1}\orcidID{0000-0001-5453-9763} \and
El\.{z}bieta~Sroka\inst{1,2}\orcidID{0000-0001-8952-2187} \and
Grzegorz~J.~Nalepa\inst{1}\orcidID{0000-0002-8182-4225}}
\authorrunning{L. do Valle Miranda et al.}
\institute{Department of Human-Centered Artificial Intelligence, Institute of Applied Computer Science, Faculty of Physics, Astronomy and Applied Computer Science, Jagiellonian University, prof. Stanis\l{}awa \L{}ojasiewicza 11, 30-348 Krak\'{o}w, Poland\\ \and 
Lukasiewicz Research Network -- Institute of Innovative Technologies EMAG, Leopolda 31, 40-189 Katowice, Poland\\
\email{luiz.dovallemiranda@doctoral.uj.edu.pl,\\\{krzysztof.kutt,grzegorz.j.nalepa\}@uj.edu.pl}}
\maketitle              %
\begin{abstract}
The rich metadata created nowadays for objects stored in libraries has nowhere to be stored, because core standards, namely MARC 21 and Dublin Core, are not flexible enough.
The aim of this paper is to summarize our work-in-progress on tackling this problem in research on cultural heritage objects at the Jagiellonian University (JU).
We compared the objects' metadata currently being collected at the JU (with examples of manuscript, placard, and obituary) with five widespread metadata standards used by the cultural heritage community: Dublin Core, EAD, MODS, EDM and Digital Scriptorium.
Our preliminary results showed that mapping between them is indeed problematic, but we identified requirements that should be followed in further work on the JU cultural heritage metadata schema in order to achieve maximum interoperability.
As we move forward, based on the successive versions of the conceptual model, we will conduct experiments to validate the practical feasibility of these mappings and the degree to which the proposed model will actually enable integration with data in these various metadata formats.

\end{abstract}

\section{Introduction and Motivation}
\label{sec:intro}

Libraries have always tried to keep up with developments in information and communication technology, especially in the search for answers to the problem of interoperability~\cite{arms2012dl}.
In the 1960s, with the spread of magnetic tapes, the MARC (Machine-Readable Cataloging) format was created, allowing for the exchange of catalog records and, after decades of work by librarians, enabling the conversion of card catalogs to electronic form.
Then, in the 1990s, as a response to the development of the World Wide Web, the Dublin Core (DC) standard emerged as a lightweight way to create metadata for indexing web pages~\cite{arms2012dl}.

The challenge is that today, 60 years after the creation of MARC and 30 years after the creation of DC, these standards are still the basis for cataloging systems and digital libraries, respectively~\cite{arms2012dl}. 
In fact, the standards have been updated (e.g., appearance of MARC 21 in 1999, and its 40th update published in 2025~\cite{marc2025bib}), but the basis of technology has remained the same and no longer fits modern IT solutions~\cite{arms2012dl,tennant2002marc}.
Despite calls for action, dating back to the 2000s, including the need to abandon existing technologies in favor of new solutions~\cite{ross2008library,tennant2002marc}, phasing out legacy technologies is difficult due to widespread adoption by libraries around the world and resistance from the community~\cite{arms2012dl}.
As a result, the rich metadata created nowadays for objects stored in libraries---both manually by researchers working in digital humanities and automatically using artificial intelligence methods---has nowhere to be stored, because these standards are not flexible.
Furthermore, these data cannot be easily integrated with other data, as the standards are only applied in a library context.

The aim of this paper is to summarize our work-in-progress on tackling this problem in research on cultural heritage objects at the Jagiellonian University (JU).
Our solution is based on linked data (LD) technology, which is not a new idea~\cite{almaaitah_opportunities_2020,Haslhofer2018,hyvonen2019knowledge,schreur2020ld}, as the required flexibility and interoperability are inherent features of its philosophy.
However, as a recent survey shows~\cite{lvm2025swjournal}, 20\% of LD-based portals in the CH domain use non-interoperable vocabularies, making the resulting datasets difficult to share.
Therefore, in our work, using the metadata available at JU, we want to demonstrate how we first identify needs (key properties), then review available domain vocabularies, and finally verify coverage of our requirements, striving for maximum reuse.

The rest of the paper is structured as follows.
In Sect.~\ref{sec:problem}, we present in detail the problem and metadata collected at JU.
An overview of selected domain vocabularies can be found in Sect.~\ref{sec:standards}, while verification of their compatibility with JU data is provided in Sect.~\ref{sec:tkm}.
Sect.~\ref{sec:summary} concludes the paper.

\section{Problem Description}
\label{sec:problem}

We have seen the need for flexible and interoperable metadata schemes firsthand while working---as a group of computer science experts---with different cultural heritage initiatives at Jagiellonian University,
like the CHExRISH project\footnote{\url{https://chexrish.id.uj.edu.pl/}}, in which three Jagiellonian University unit---namely the Jagiellonian University Museum, the Jagiellonian University Archives, and the Jagiellonian Library---are digitizing, describing, and interlinking collections of archival documents related to the university's history.
Researchers working with these holdings create rich descriptive metadata~\cite{hussain2005metadata,nahotko2004metadane} beyond the capabilities of MARC 21 and DC standards (see Tab.~\ref{tab:data}).

\begin{table}[pth]
    \caption{Actual description of sample manuscript (not digitzed yet), placard~\cite{TKMcyrkSidoli} and obituary~\cite{TKMobituary} from JU research teams.}\label{tab:data}
    \centering
    \scriptsize
    \begin{tabular}{p{1.7cm}p{3.35cm}p{3.35cm}p{3.35cm}}
    \hline
    \textbf{Property} & \textbf{Manuscript} & \textbf{Placard} & \textbf{Obituary} \\
    \hline \hline
    
    \emph{Title} & %
    Letter from [Karl] Heinrich Aster to [Johann David Erdmann] Preuss & %
    [incipit] Circus Sidoli near the Castle 4th guest performance of the famous athlete Mr. Holtum & %
    [incipit] Monsieur et Madame Victor Chodźko et leurs enfants, [...]  \\ %
    
    \emph{Alternative title} & %
    -- & %
    Sidoli Circus under the Castle 4 guest performance by the famous athlete Mr. Holtum & %
    Headline title.
On the hourglass stated: Administration des Funérailles. PAUL HALL, Directeur, J. Pl. St. Suspice et 1, R. Férou. \\ %
    
    \emph{Author} & %
    [Karl] Heinrich Aster & %
    NN & %
    NN \\ %
    
    \emph{Type of document} & %
    Letter & %
    Placard & %
    Obituary; Hourglass; Leaflet \\ %
    
    \emph{Language} & %
    German & %
    Polish & %
    French \\ %
    
    \emph{Identifier} & %
    SA, Aster, Karl Heinrich\_2.2 & %
    NDIGDZS033452 & %
    NDIGDZS062327 \\ %
    
    \emph{Physical extent} & %
    4 pages & %
    1 card; 57x88 cm & %
    4 pages; 27 cm \\ %
    
    \emph{Material information} & %
    Red seal and postage stamp: "Sachsen. Neu 3 Grosch" (card 5v) & %
    -- & %
    Black stamp and handwriting [page 3] \\ %
    
    \emph{Place of origin} & %
    Dresden & %
    Cracow & %
    Paris \\ %
    
    \emph{Creation Date} & %
    28.12.1852 & %
    1878 & %
    1891 \\ %
    
    \emph{Current location} & %
    Cracow & %
    Cracow & %
    Cracow \\ %
    
    \emph{Custody history} & %
    Radowitz 5262 & %
    -- & %
    -- \\ %
    
    \emph{Publisher} & %
    -- & %
    -- & %
    -- \\ %
    
    \emph{Other editions} & %
    -- & %
    -- & %
    -- \\ %
    
    \emph{Related date} & %
    -- & %
    Event: 24.01.1878 & %
    Date of death: 19.12.1891; Date of funeral: 23.12.1891 \\ %
    
    \emph{Related place} & %
    Sender location: Dresden & %
    Event place: Cracow & %
    Les Champeaux Cemetery (Montmorency); Church of the Virgin Mary of the Assumption (Paris) \\ %
    
    \emph{Related person} & %
    Sender: Aster, Karl Heinrich; Receiver: Preuss, Johann David Erdmann & %
    Artist: John (?) Holtum; Artist: Oskar Mink & %
    Deceased: Chodźko, Alexander (1804-1891); Mentioned: Victor Chodźko  \\ %
    
    \emph{External link} & %
    -- & %
    \url{https://jbc.bj.uj.edu.pl/publication/602279} & %
    \url{https://jbc.bj.uj.edu.pl/publication/860874} \\ %
    
    \emph{Description / Notes} & %
    -- & %
    In the document description of attractions; prices of places; text in several languages appears  & %
    On the hourglass, the number of years of life is incorrect. It should read: en sa 87..e année \\ %
    
    \emph{Typography note} & %
    -- & %
    -- & %
    Italic font; No. 13 in upper right corner, large M in center  \\ %
    
    \emph{Keywords} & %
    19th century & %
    ephemeral prints; 19th century & %
    ephemeral prints; 19th century; orientalists; funeral; professors; religion and spirituality; society \\ %
    \hline
    \end{tabular}
\end{table}

One solution could be to place excess metadata in existing fields (e.g., unused MARC 21 fields), based on internal institutional policies, but this leads to a situation where data is not interoperable, as different institutions may have different information in the same field.
The second solution would be to use a free-text field (description in DC, field 59X in MARC 21), but this leads to a situation where one field serves many purposes~\cite{burke2020descriptive}, which hinders its automatic processing and interoperability.
The third idea is to place additional information as keywords (e.g.~\cite{TKMobituary}), but in this solution the meaning of individual words is lost (whether they denote the type of document, the place of the event, the person depicted in the picture, etc.).
Storing this data in Notepad, Word, or PDF files is also not a good solution, as it prevents interoperability.

As many authors suggest, e.g.~\cite{almaaitah_opportunities_2020,Haslhofer2018,hyvonen2019knowledge,schreur2020ld}, the use of LD seems to be a response to indicated problem and a key element in the transformation of library services~\cite{schreur2020ld}.
LD is based on the endeavor to represent data using shared vocabularies for both individuals, their characteristics, and the relation between individuals.
Such a metadata vocabulary references unique web identifiers (URI) in a machine-readable way that allows linkage to semantically related resources from other datasets (for an introduction to LD technology in a library context, see, e.g.:~\cite{papadakis2015ld,subirats2020lodebd3}).
Modeling a certain dataset using LD facilitates data integration between institutions and certain operations with the dataset, including knowledge discovery, enrichment, and visualization~\cite{hyvonen2019knowledge}.
Therefore, LD directly addresses one of the main issues related to the use of information and communication technologies in libraries mentioned above, namely interoperability~\cite{almaaitah_opportunities_2020}.
However, the question arises: how to choose the right shared vocabulary for project needs?

\section{Review of Selected Vocabularies}
\label{sec:standards}

First of all, one needs to review the existing shared vocabularies related to the project needs.
As shown in Tab.~\ref{tab:data}, our needs include descriptive metadata (excluding administrative, technical, and other metadata~\cite{nahotko2004metadane}) in the area of cultural heritage.
The popular vocabulary within this scope includes~\cite{gaitanou2024ld,ranjgar2024ch,silva2024ch,subirats2020lodebd3,lvm2025swjournal}:

\begin{enumerate}

\item Generic vocabularies:
\begin{itemize}
\item \emph{Dublin Core} (\url{https://www.dublincore.org/}) -- one of the more generic vocabularies for Linked Data.

\item \emph{FOAF} (Friend of a Friend; \url{http://xmlns.com/foaf/spec/}) -- a %
scheme used to describe social relationships between individuals.

\item \emph{SKOS} (Simple Knowledge Organization System; \url{https://www.w3.org/2004/02/skos/}) -- %
a standard data model for sharing and linking knowledge organization systems via the semantic web.

\item \emph{\url{https://schema.org}}---particularly the CreativeWork and derived classes---used to mark the most important information on web pages for more accurate indexing by search engines.
\end{itemize}

\item Standards for describing library resources, incl. attempts to create a modern cataloguing system to replace MARC 21~\cite{gaitanou2024ld}:
\emph{MODS} (\url{https://www.loc.gov/standards/mods/}),
\emph{FRBR} (\url{https://www.loc.gov/catdir/cpso/frbreng.pdf}),
\emph{BIBFRAME} (\url{https://www.loc.gov/bibframe/}),
\emph{RDA} (Resource Description and Access; \url{https://www.loc.gov/aba/rda/mgd/}),
\emph{BIBO} (\url{https://dcmi.github.io/bibo/}).

\item Standards used to describe archival material:
\begin{itemize}
\item \emph{EAD}, an international XML-based standard %
used, e.g., in Kalliope Union Catalog founded by the Berlin State Library\footnote{\url{https://kalliope-verbund.info/en/index.html}}. 

\item \emph{Records in Contexts Ontology} (RiC-O; \url{https://www.ica.org/resource/records-in-contexts-ontology/}), which enables the representation of archival information within a broader context~\cite{ric2017primer}.
\end{itemize}

\item Event-based schemes (in contrast to earlier ones, which were document-based):
\begin{itemize}
\item \emph{LIDO} (\url{https://lido-schema.org/}), designed to store descriptions of cultural heritage objects, related events, and administrative information.

\item \emph{CIDOC-CRM} (\url{https://cidoc-crm.org/}), an international standard (ISO 21127:2014) for cultural heritage data modeling and sharing across different institutions~\cite{cidoc2014primer},

\item \emph{LRMoo} (\url{https://cidoc-crm.org/lrmoo}; previously known as \emph{FBRoo}), one of the extensions of CIDOC-CRM,
\end{itemize}

\item Metadata schemes created by and for aggregators:
\begin{itemize}
\item \emph{Europeana Data Model} (EDM; \url{https://pro.europeana.eu/page/edm-documentation}) -- a model for describing the metadata of various objects of GLAM sector institutions stored in the Europeana system built, among others, on the Dublin Core and LIDO.

\item \emph{Digital Scriptorium 2.0} (DS; \url{https://digital-scriptorium.org/}) -- metadata schema and %
a semantic portal and knowledge base for manuscripts from multiple sources in a single interface. %
\end{itemize}

\item Content (e.g., information about specific paragraphs) and structure (e.g., information about specific files) description:
\begin{itemize}
\item \emph{METS} (\url{https://mets.github.io/}) -- 
a structured format (container) for organizing, structuring and managing digital objects. 

\item \emph{TEI-XML} (\url{https://tei-c.org/}) and \emph{EpiDoc} (\url{https://epidoc.stoa.org/}) -- international standards for editing and publishing digital editions.

\item \emph{ALTO} (\url{https://www.loc.gov/standards/alto/}) %
-- an XML format for storing layout and OCR-recognized text.
\end{itemize}

\item Thesauri:

\begin{itemize}
\item \emph{ICONCLASS} (\url{http://www.iconclass.org/}) for the detailed classification of the content of graphic objects.

\item \emph{Getty AAT} (\url{https://www.getty.edu/research/tools/vocabularies/aat/})  -- specialized terminology for art and architecture.

\item \emph{GeoNames} (\url{https://www.geonames.org/}) -- basic geographical vocabulary in Linked Data.
\end{itemize}

\end{enumerate}

\section{Comparison of Metadata Schemes}
\label{sec:tkm}

Despite the differences between vocabularies, propositions exist on how to map elements from one standard to another, including mappings from EAD to EDM~\cite{hennicke2011conversion,rockenbauer2012deliverable} and CIDOC-CRM~\cite{theodoridou2001mapping}, from RiC-O to CIDOC-CRM~\cite{Bountouri2023},
and between FRBR, BIBFRAME and RDA~\cite{gaitanou2024ld}.
Drawing on the advancements presented by such works, and by our own familiarization with specifications, Tab.~\ref{tab:comparison} compares list of properties from Tab.~\ref{tab:data}, Dublin Core, EAD, MODS, EDM and Digital Scriptorium as a guideline for the development of a standard for representing metadata for CH artifacts stored at JU.
These five were selected because they are basic description standards (Dublin Core, MODS), are used in related collections (for historical reasons, some collections are divided between the Jagiellonian Library and the Berlin State Library, so for interoperability, the EAD standard used by the latter should be considered) and their use in aggregators, which will allow for easier integration with the resources they store (EDM, Digital Scriptorium).

\begin{table}[pth]
    \caption{Comparison of the forming standard for describing metadata of cultural heritage objects at JU (column \emph{Property}) with Dublin Core, EAD, MODS, EDM and Digital Scriptorium.}\label{tab:comparison}
    \centering
\rotatebox{90}{
    \scriptsize
\begin{tabular}{p{1.8cm}p{3.1cm}p{3.1cm}p{3.1cm}p{3.3cm}p{2.9cm}}
\hline
\textbf{Property} & \textbf{DublinCore} & \textbf{EAD} & \textbf{MODS} & \textbf{EDM} & \textbf{Digital Scriptorium} \\
\hline \hline
\emph{Title} & \lstinline{dc:title} & \lstinline{<unittitle>} & \lstinline{<titleInfo>} & \lstinline{dc:title} & \lstinline{Title} \\
\emph{Alternative title} & \lstinline{dcterms:alternative} & \lstinline{--} & \lstinline{<titleInfo type="alternative">} & \lstinline{dcterms:alternative} & \lstinline{--} \\
\emph{Author} & \lstinline{dc:creator} & \lstinline{<persname> or <corpname>} & \lstinline{<name>} & \lstinline{dc:creator} & \lstinline{Author} \\
\emph{Type of document} & \lstinline{dc:type} & \lstinline{<controlaccess><genreform></controlaccess>} & \lstinline{<typeOfResource>} & \lstinline{dc:type or edm:hasType or edm:type} & \lstinline{--} \\
\emph{Language} & \lstinline{dc:language} & \lstinline{<langmaterial><language></langmaterial>} & \lstinline{<language>} & \lstinline{dc:language} & \lstinline{Language} \\
\emph{Identifier} & \lstinline{dc:identifier} & \lstinline{<unitid>} & \lstinline{<identifier>} & \lstinline{dc:identifier} & \lstinline{Shelfmark} \\
\emph{Physical extent} & \lstinline{format} & \lstinline{<physdesc><extent></physdesc>} & \lstinline{<physicalDescription>} & \lstinline{dcterms:extent} & \lstinline{Physical Description} \\
\emph{Material information} & \lstinline{--} & \lstinline{<physdesc>} & \lstinline{--} & \lstinline{--} & \lstinline{--} \\
\emph{Place of origin} & \lstinline{--} & \lstinline{<geogname>} & \lstinline{<orginInfo><place></origInfo>} & \lstinline{edm:hasMet} & \lstinline{Place} \\
\emph{Creation Date} & \lstinline{dcterms:created} & \lstinline{<unitdate>} & \lstinline{<orginInfo><dateCreated></origInfo>} & \lstinline{dcterms:created} & \lstinline{Date} \\
\emph{Current location} & \lstinline{--} & \lstinline{<physloc>} & \lstinline{<orginInfo><place></orginInfo>} & \lstinline{edm:currentLocation} & \lstinline{Holding Institution} \\
\emph{Custody history} & \lstinline{dcterms:provenance} & \lstinline{<custodhist>} & \lstinline{<orginInfo>} & \lstinline{dcterms:provenance} & \lstinline{--} \\
\emph{Publisher} & \lstinline{dc:publisher} & \lstinline{<bibref>} & \lstinline{<orginInfo><publisher></orginInfo>} & \lstinline{dc:publisher} & \lstinline{--} \\
\emph{Other editions} & \lstinline{dcterms:hasVersion or dcterms:isVersionOf} & \lstinline{<bibliography>} & \lstinline{<orginInfo><edition></orginInfo>} & \lstinline{dcterms:hasVersion or dcterms:isVersionOf} & \lstinline{--} \\
\emph{Related date} & \lstinline{dcterms:temporal} & \lstinline{--} & \lstinline{<subject><temporal></subject>} & \lstinline{edm:hasMet} & \lstinline{--} \\
\emph{Related place} & \lstinline{dcterms:spatial} & \lstinline{<geogname>} & \lstinline{--} & \lstinline{dcterms:spatial or edm:happenedAt} & \lstinline{--} \\
\emph{Related person} & \lstinline{dc:contributor or dc:creator} & \lstinline{<persname>} & \lstinline{--} & \lstinline{dc:contributor or dc:creator or edm:hasMet or edm:isRepresentationOf} & \lstinline{--} \\
\emph{External link} & \lstinline{dc:relation} & \lstinline{--} & \lstinline{<location> <url>} & \lstinline{dc:relation or edm:isRelatedTo} & \lstinline{Institutional Record} \\
\emph{Description / Notes} & \lstinline{dc:description} & \lstinline{<scopecontent>} & \lstinline{<abstract> or <note> or <tabelofContents>} & \lstinline{dc:description} & \lstinline{Note} \\
\emph{Typography note} & \lstinline{--} & \lstinline{--} & \lstinline{--} & \lstinline{--} & \lstinline{--} \\
\emph{Keywords} & \lstinline{dc:subject} & \lstinline{<controlaccess><subject></controlaccess>} & \lstinline{<subject> or <classification>} & \lstinline{dc:subject} & \lstinline{Keyword(s)} \\
\hline
\end{tabular}
}
\end{table}

The first column of Tab.~\ref{tab:comparison} presents the concept from our current work-in-progress model, while five others present related elements from selected standards.
Some elements are represented in the similar way in all standards, for example, ``Title'', ``Description / Notes'', ``Physical Extent'', ``Identifier'', ``Creation date'', and ``Language''.

Another case of mapping is when the representations are slightly different, but can be unproblematically mapped, such as (1) ``Place of origin'' and (2) ``Related person''. In the case of (1), EDM has no element for encoding the place of origin, the place is connected with the object by the affirmation that the object has merely been in this place\footnote{See~\cite{peroni2013} for a more extensive highlighting of the limitations of the EDM.}. Regarding (2), EAD encodes it as a reference to a person, while EDM as a representation or potential meeting (\lstinline[basicstyle=\ttfamily\footnotesize]{edm:hasMet}).

The last set of concepts is those whose mapping cannot be made one to one, but need multiple elements from one standard to be mapped to multiple elements from the other. The most important of those concepts is ``Related person''. This concept includes subconcepts such as ``Creator'', ``Sender'', ``Receiver'', ``Issuer'', etc.  One clear example of such a difficulty is the pair ``Sender'' and ``Receiver''. While in Kalliope these roles can be easily implemented, in EDM they are problematic.
One solution is to represent the sender of a letter as the creator of the letter. However, it is worth noting that it is an approximation, as it is not necessarily the same as referring to the creator of an academic work.
The receiver can be considered as a mere person to whom the letter has come in contact (\lstinline[basicstyle=\ttfamily\footnotesize]{edm:hasMet}), without acknowledging the intentionality of the letter. 

Finally, it is worth highlighting that the set of elements present in the comparison table is not exhaustive. The rationale for constructing such a table is to make it agnostic to the content of the manuscripts, to avoid the need for a thorough study of the object before collecting metadata. However, the table still fulfills the minimum requirements for the highest level of integration to both Kalliope and Europeana and provides rich metadata for computationally aided research in the field of cultural heritage.

\section{Conclusion and Future Works}
\label{sec:summary}

In this work, we compared the objects metadata currently being collected at the JU (with examples of manuscript, placard, and obituary) with five widespread metadata standards: Dublin Core, EAD, MODS, EDM and Digital Scriptorium.
This comparison indicates that the very mapping of metadata %
is indeed problematic, and that the following requirements should be followed in the development of a conceptual model for JU cultural heritage metadata.
A conceptual model to be adopted for JU's metadata, thus, needs to be nuanced enough to be interoperable with the more abstract representation of EDM and more granular classification allowed by EAD.
Moreover, the model should be able to expliclty differentiate between the physical resource and its digital counterpart, somewhat following EDM's distinctions.
At last, the model should be flexible enough to accommodate further metadata coming from other subunits of the JU.

With maximum interoperability in mind, we want the model that will be created to enable the mappings we have determined to the largest extent possible. As we move forward, based on the successive versions of the conceptual model, we will conduct experiments to validate the practical feasibility of these mappings and the degree to which the proposed model will actually enable integration with data in these various metadata formats.

\begin{credits}
\subsubsection{\ackname}
This publication was funded by a flagship project ``CHExRISH: Cultural Heritage Exploration and Retrieval with Intelligent Systems at Jagiellonian University'' under the Strategic Programme Excellence Initiative at Jagiellonian University.
The research has been supported by a grant from the Priority Research Area (DigiWorld) under the Strategic Programme Excellence Initiative at Jagiellonian University.

This work benefited from the use of language models to support proofreading and enhance readability.

\subsubsection{\discintname}
The authors have no competing interests to declare that are
relevant to the content of this article.
\end{credits}

\bibliographystyle{splncs04}
\bibliography{geistbib/culheripub,geistbib/culheriteam}

\end{document}